\documentclass[11pt]{article}
\pdfoutput=1
\usepackage[affil-it]{authblk}
\usepackage{geometry}               
\usepackage{epstopdf}
\usepackage{fullpage}
\usepackage[super,sort&compress,comma]{natbib}

\usepackage{array}
\usepackage{caption}
\usepackage{float}
\usepackage{natbib}
\usepackage{setspace}
\usepackage{xkeyval}
\usepackage{amsmath,color}
\usepackage{graphicx}
\usepackage{bm}
\usepackage{color}
\usepackage{verbatim}
\usepackage{amssymb}
\usepackage{multirow}
\usepackage{mathtools}
\usepackage{bbold}
\usepackage[dvipsnames]{xcolor}
\usepackage{soul}
\usepackage{xcolor}
\usepackage{transparent}

\newcommand{\bra}[1]{\langle #1|}
\newcommand{\ket}[1]{|#1\rangle}

\newcommand{\rvec}{\mathbf{R}}

\newcommand{\Pvec}{\mathbf{P}}

\newcommand{\proj}{\mathbb{P}}

\newcommand{\Qvec}{\mathbf{Q}}

\title{State Space Path Integrals for Electronically Nonadiabatic Reaction Rates}
\author{Jessica R. Duke and Nandini Ananth\thanks{Email: ananth@cornell.edu}}
\affil{Department of Chemistry and Chemical Biology, Cornell University, Ithaca, New York 14853, USA}

\begin{document}
\newcommand{\cev}[1]{\reflectbox{\ensuremath{\vec{\reflectbox{\ensuremath{#1}}}}}}

\maketitle

\begin{abstract}
We present a state-space-based path integral 
method to calculate
the rate of electron transfer (ET) 
in multi-state, multi-electron condensed-phase
processes. 
We employ an exact path integral in discrete 
electronic states and continuous Cartesian nuclear
variables to obtain a transition state theory (TST)
estimate to the rate. A dynamic recrossing correction to the 
TST rate is then obtained from real-time dynamics
simulations using mean field ring polymer molecular 
dynamics. 
We employ two different reaction coordinates in our 
simulations 
and show that, despite the use of mean field 
dynamics, the use 
of an accurate dividing surface to compute TST 
rates allows us to achieve remarkable agreement with 
Fermi's golden rule rates for nonadiabatic ET 
in the normal regime of Marcus theory.
Further, we show that using 
a reaction coordinate based on electronic state
populations allows us to capture the turnover 
in rates for ET in the Marcus inverted regime.
\end{abstract}

\section{Introduction}
Condensed-phase electron transfer (ET) reactions drive a wide range of
energy conversion and catalytic pathways in 
biological systems\cite{Marcus1985,Gray1996,Reece2009} and renewable energy devices.\cite{Cukier1998,Lewis2006,Feldt2013}
Developing theoretical methods capable of accurately calculating 
rate constants for these reactions is an ongoing challenge
and a crucial step towards the design of materials with desirable charge 
and energy transfer properties.
While numerous mixed quantum-classical~\cite{Tully1990,Kapral2006,Jain2015} 
and 
semiclassical methods\cite{Cao1995,Cao1997,
Ananth2007,Cotton2013,Lee2016}
for simulating 
ET reactions in the condensed phase have been developed over the years, 
they are limited by either computational complexity or the use of 
dynamics that fail to preserve detailed balance.
Alternatively, methods based on imaginary-time path integrals (PIs) such as 
centroid molecular dynamics (CMD)\cite{Cao1994} and ring 
polymer molecular dynamics (RPMD) \cite{Craig2004}
that employ classical trajectories to capture quantum dynamics 
effects have emerged as promising methods for the computation
of condensed-phase reaction rates.\cite{Craig2005,Hele2013,Hele20132,Hele2016}
RPMD in particular has been successfully employed to study a variety 
of chemical reactions\cite{Menzeleev2011,Habershon2013,Wilkins2015,
Kowalczyk2015,Kretchmer2016}
and was shown to accurately predict thermal rate constants for ET
in the normal and activationless regimes of Marcus theory.\cite{Menzeleev2011}
More recently, extensions of RPMD to systems with multiple coupled 
electronic states have been developed;\cite{Richardson2013,Ananth2013,Menzeleev2014,Duke2015}
notably, the kinetically constrained (KC)-RPMD method\cite{Menzeleev2014}
accurately describes the ET reactions of two-level systems both in 
the normal and inverted regimes of Marcus theory.

In this paper, we present a simple and accurate method to calculate
rate constants for nonadiabatic ET reactions. We first 
evaluate the transition state theory (TST) rate estimate using an exact
state space path integral (SS-PI) to compute the probability of reaching
the transition state (dividing surface) from the reactant state.
The dynamic recrossing factor to correct the TST rate is then 
computed using mean field (MF)-RPMD,\cite{MF1}
with trajectories initialized to the dividing surface.
This approach generalizes the standard RPMD implementation to 
multi-state, multi-electron systems with very little additional complexity
and retains all the desirable features of RPMD including, most notably,
the conservation of detailed balance. We obtain quantitatively accurate
rates in the normal regime of ET using two different reaction coordinates,
and we capture the qualitative rate turnover in the inverted regime.
Despite the use of MF-RPMD, the choice of TST dividing surface
allows us to obtain numerically accurate reaction 
rates for ET over the full range of electronic coupling strengths 
spanning six orders of magnitude.

This paper is organized as follows:
In Section 2 we review
general reaction rate theory, 
the state space path integral discretization of 
the quantum partition function, and the MF-RPMD formulation.
In Section 3, we discuss our approach to reaction rates 
for multi-state systems and introduce the different 
reaction coordinates. 
We present the details of the simulation used to obtain
the TST rate estimate and the details of the MF-RPMD simulation used 
to obtain the dynamic recrossing factor in Section 4. 
In Section 5, we specify the model systems employed here that
explore a range of driving forces and electronic coupling strengths.
Finally, we discuss our results in Section 6 and conclude in Section 7.

\section{Methods}
\subsection{Reaction Rate Theory}
We start by reviewing the general theory of reaction rates and introduce 
the specific formulation relevant to our simulation protocol.
As with other RPMD-based methods, our SS-PI formulation 
allows us to exploit standard techniques for calculating 
classical reaction rates.
The reaction rate constant
can be written in terms of a flux-side correlation function, 
\cite{Miller1983,Chandler1987}
\begin{equation}
	k=
	\lim_{t \to \infty}
	\frac{\left\langle 
	\delta\left(\xi\left(\mathbf{y}_0\right) - \xi^\ddag\right)
	\dot \xi_0 \;
	h \left(\xi\left(\mathbf{y}_t\right) - \xi^\ddag \right) 
	\right\rangle}
	{\left\langle 
	h \left(\xi^\ddag - \xi\left(\mathbf{y}_0\right)\right)
	\right\rangle},
	\label{eq:singlek}
\end{equation}
where the angular brackets indicate canonical ensemble averages,
$h$ represents the Heaviside function, and $\delta$ is the Dirac delta function.
In Eq.~\eqref{eq:singlek}, we use a general reaction coordinate, 
$\xi (\mathbf{y})$, that is a function of nuclear and electronic state
variables, $\mathbf{y}=\left\{ \rvec, n \right\}$,
and that
distinguishes between reactants and products via 
the dividing surface defined as $\xi (\mathbf{y}) = \xi^\ddag$.
Throughout, we use bold notation to indicate multi-dimensional vectors.
Following the Bennett-Chandler approach,\cite{Frenkel2002}
Eq.~\eqref{eq:singlek} can be factored into a purely statistical TST rate estimate, 
$k_\text{TST}$, 
and a time-dependent coefficient,  
$\kappa(t)$, that accounts for dynamic recrossing at the dividing surface:
\begin{equation}
	k=
	\frac{\left\langle 
	\dot \xi_0 \; h\left(\dot \xi_0 \right) \right\rangle_c
	\left\langle
	\delta\left(\xi\left(\mathbf{y}_0\right) - \xi^\ddag\right)
	\right\rangle}
	{\left\langle 
	h \left(\xi^\ddag - \xi\left(\mathbf{y}_0\right)\right)
	\right\rangle}
	\times
	\lim_{t \to \infty}
	\frac{\left\langle 
	\delta\left(\xi\left(\mathbf{y}_0\right) - \xi^\ddag\right)
	\dot \xi_0 \;
	h\left(\xi
	\left( \mathbf{y}_t \right)
	 - \xi^\ddag \right)
	\right\rangle}
	{\left\langle 
	\dot \xi_0 \; h\left(\dot \xi_0 \right) 
	\right\rangle_c
	\left\langle
	\delta\left(\xi\left(\mathbf{y}_0\right) - \xi^\ddag\right)
	\right\rangle},
	\label{eq:splitknuc}
\end{equation}
where $\left\langle\cdots\right\rangle_c$ indicates an ensemble
average with the system constrained to the TS corresponding to a particular reaction coordinate.

\subsection{State Space Path Integral Discretization}
Next, we review the imaginary-time SS-PI discretization used to obtain the TST rate estimate.
Consider the Hamiltonian for a general 
$K$-level system with $d$ nuclear degrees of freedom (dofs) in the diabatic representation:
\begin{equation}
	\label{eq:hamiltonian}
	\hat{H} = 
	\sum_{j=1}^d \frac{\hat{P}_j^2}{2 M_j} + 
	\sum_{n,m=1}^K \ket{n} V_{nm} (\hat{\rvec}) \bra{m},
\end{equation}
where $\hat{\rvec}$ and $\hat{\Pvec}$
represent nuclear position and momentum operators, 
respectively, $M$ is nuclear mass, \{$\ket{n}$\} are diabatic 
electronic states, and \{$V_{nm}(\rvec)$\} are diabatic potential 
energy matrix elements.
PI discretization of
the quantum canonical partition function
in the product space of diabatic electronic states and nuclear position gives
\begin{eqnarray}
	\label{eq:pfdisc}
	Z = \text{Tr}\left[ e^{-\beta \hat{H}} \right] =
	 \int \left\{d\rvec_\alpha \right\} 
	\sum_{\left\{n_\alpha \right\} =1}^K
	\prod_{\alpha=1}^{N}
	\bra{\rvec_\alpha,n_\alpha} e^{-\frac{\beta}{N} 
	\hat{H}} \ket{\rvec_{\alpha+1},n_{\alpha+1}},
\end{eqnarray}
where $\beta\!~=~\!1/k_BT$, $T$ is temperature, $N$ is the number of imaginary time slices or ``beads,''
$(\rvec_\alpha,n_\alpha)$ refers to the nuclear position and electronic state of the $\alpha^{th}$ bead, 
$(\rvec_{N+1},n_{N+1})  = (\rvec_1,n_1)$, and 
we use the notations
$\int\left\{d\rvec_\alpha \right\}= \int d\rvec_1\int d\rvec_2 \ldots\int d\rvec_N$ and 
$\sum_{\left\{n_\alpha\right\}=1}^K=\sum_{n_1=1}^K\sum_{n_2=1}^K\ldots\sum_{n_N=1}^K$.

Applying the standard short-time approximations\cite{Trotter1959,Chandler1987} 
to evaluate the matrix elements in Eq.~\eqref{eq:pfdisc} and setting $\hbar = 1$,
we obtain the expression
\begin{eqnarray}
	\label{eq:pfss}
	Z \propto \lim_{N\to\infty} \int \left\{d\rvec_\alpha \right\} 
	e^{-\frac{\beta}{N} 
	V_N
	\left( \{\rvec_\alpha\}\right)}
	\text{Tr} \left[ \Gamma \right],
\end{eqnarray}
where the proportionality sign indicates pre-multiplicative constants 
have been omitted for simplicity.
In Eq.~\eqref{eq:pfss},
\begin{eqnarray}
	\label{eq:rppot}
	V_N =
	\sum_{j=1}^d
	\sum_{\alpha=1}^N \left[ \frac{M_j N^2}{2\beta^2}
	(R_{j,\alpha}-R_{j,\alpha+1})^2 \right],
\end{eqnarray}
\begin{equation}
	\label{eq:gamma}
	\Gamma = 
	\prod_{\alpha=1}^{N} \textbf{M} (\rvec_\alpha),
\end{equation}
and the $K\times K$-dimensional interaction matrix \textbf{M} has elements
\begin{equation}
	\textbf{M}_{n_\alpha n_{\alpha+1}} (\rvec_\alpha) = 
	\begin{cases}
		e^{-\frac{\beta}{N} V_{n_\alpha n_{\alpha}}(\rvec_\alpha)} & n_\alpha=n_{\alpha+1} \\
		-\frac{\beta}{N} V_{n_\alpha n_{\alpha+1}}(\rvec_\alpha)\;
		e^{-\frac{\beta}{N} V_{n_\alpha n_{\alpha}}(\rvec_\alpha)} & n_\alpha \ne n_{\alpha+1}.\\
	\end{cases}
	\label{eq:mmatrix}
\end{equation}
We note that the trace of $\Gamma$ will be, in general, positive for all 
$K$-level systems when the off-diagonal diabatic coupling matrix elements
are positive.

The canonical ensemble average of an observable in the SS-PI framework 
can be written as
\begin{equation}
	\left\langle \hat{A}\right\rangle = 
	\frac{1}{Z}\text{Tr}[e^{-\beta \hat{H}} A(\hat{\rvec})]
	=\frac{
	\int \left\{d\rvec_\alpha \right\} 
	e^{-\frac{\beta}{N} V_N
	\left( \{\rvec_\alpha\}\right)}
	\text{Tr} \left[ \Gamma \right]
	A\left(\{\rvec_\alpha\}\right)
	}
	{
	\int \left\{d\rvec_\alpha \right\} 
	e^{-\frac{\beta}{N} 
	V_N
	\left( \{\rvec_\alpha\}\right)}
	\text{Tr} \left[ \Gamma \right]
	}
	\label{eq:can_avg}
\end{equation}
and can be evaluated using standard Monte Carlo (PIMC) 
or molecular dynamics (PIMD) methods that converge 
to the exact result in the limit $N\rightarrow\infty$.

\subsection{Mean Field RPMD}
The dynamic recrossing factor (second term in Eq.~\eqref{eq:splitknuc}) 
is calculated using MF-RPMD, briefly reviewed here.
Exponentiating the trace in Eq.~\eqref{eq:pfss} and multiplying
by normalized Gaussian momentum integrals for the nuclear degrees of freedom 
allows us to write the quantum partition function
in terms of a classical ring polymer Hamiltonian:
\begin{eqnarray}
	\label{eq:pfmf}
	Z \propto \lim_{N\to\infty} \int \left\{d\rvec_\alpha \right\} 
	\int \left\{d\Pvec_\alpha \right\} 
	e^{-\frac{\beta}{N} 
	H_N \left( \{\rvec_\alpha\}, 
	\{\Pvec_\alpha\} \right)},
\end{eqnarray}
where
\begin{eqnarray}
	\label{eq:mfham}
	H_N =
	\sum_{j=1}^d
	\sum_{\alpha=1}^N \left[ \frac{M_j N^2}{2\beta^2}
	(R_{j,\alpha}-R_{j,\alpha+1})^2
	+ \frac{P_{j,\alpha}^2}{2M_j} \right] 
	- \frac{N}{\beta}\text{ln}\left(  
	\text{Tr} \left[ \Gamma \right] \right).
\end{eqnarray}
The dynamic recrossing factor in the MF-RPMD framework is written as
 \begin{equation}
	\kappa_{\text{MF-RPMD}} \left( t \right) = 
	\lim_{t \to \infty}
	\lim_{N \to \infty}
	\frac{
	\left\langle
	\dot{\xi}_0 \;
	h \left(\xi_t - \xi^\ddag \right) 
	\right\rangle_c}
	{\left\langle
	\dot\xi_0 \;
	h\left(\dot\xi_0\right)
	\right\rangle_c},
	\label{eq:kappa}
\end{equation}
where $\xi\equiv\xi\left(\{\rvec_\alpha\},\{n_\alpha\} \right)$.
In Eq.~\eqref{eq:kappa}, values of the reaction coordinate at time $t$ are
obtained from classical trajectories generated by the Hamiltonian 
in Eq.~\eqref{eq:mfham}.

\section{Reaction Rate Theory Using SS-PIs and MF-RPMD}
The TST rate (the first term in Eq.~\eqref{eq:splitknuc}) 
is the product of the average
positive velocity of the reaction coordinate at the TS 
barrier and 
the probability of the system reaching 
TS configurations, $\xi^\ddag$, from its initial 
reactant state configurations.
For a system where electronic states are coupled to nuclear dofs, 
we define $\xi^\ddag$ in terms of a simultaneous restraint 
on the nuclear and electronic state configurations.
The TST rate, $k_{\text{TST}}$, can then be 
expressed as
\begin{equation}
	k_{\text{TST}} = 
	\left\langle 
	\dot \xi_0 \; h\left(\dot \xi_0 \right) \right\rangle_c \times
	\text{P} \left( \rvec^\ddag, n^\ddag\right).
	\label{eq:ktst}
\end{equation}
In this section, we discuss the definition of the transition
state dividing surface in the SS-PI representation, and 
we introduce the corresponding choice of reaction coordinate 
employed in real-time MF-RPMD simulations.
For clarity, we discuss the choice of reaction coordinate in 
the context of standard system-bath models for ET where a
multi-state system is coupled to a dissipative bath 
via a single collective solvent coordinate, but the ideas
presented here are easily generalized.

\subsection{The Solvent Coordinate}
The first reaction coordinate we employ for the MF-RPMD dynamic recrossing 
factor in Eq.~\eqref{eq:kappa} is the solvent coordinate, defined as
the center of mass (COM) of the solvent ring polymer:
$\xi \equiv \overline{\rvec}=\sum_{\alpha=1}^N \rvec_\alpha/N$.
We then define the corresponding TS as follows:
We restrain $\overline{\rvec}$ to 
the point of degeneracy between the two diabatic potential energy surfaces,
denoted by $\rvec^\ddag$.
In addition, we
limit electronic RP configurations to those where at least one bead
is in a different electronic state than the others.  With this
definition of the TS, the reactant
state is defined by electronic 
RP configurations where at least one electronic bead 
is in the reactant state and, for ET in the normal regime, 
$\overline{\rvec} \le \rvec^\ddag$.

The probability of reaching the 
TS from reactants, $\text{P} \left( \rvec^\ddag, n^\ddag \right)$
in Eq.~\eqref{eq:ktst}, is defined as
\begin{eqnarray}
	\label{eq:prdagndag}
	\text{P} \left( \rvec^\ddag, n^\ddag \right) = 
	\frac{
	\int \left\{d\rvec_\alpha \right\} 
	e^{-\frac{\beta}{N} 
	V_N
	\left( \{\rvec_\alpha\}\right)}
	\text{Tr} \left[ \Gamma_\text{Kinks}
	\right]
	\delta \left( \overline{\rvec} - \rvec^\ddag \right)
	}
	{
	\int_{-\infty}^{\rvec^\ddag} d\rvec^\prime
	\int \left\{d\rvec_\alpha \right\} 
	e^{-\frac{\beta}{N} 
	V_N
	\left( \{\rvec_\alpha\}\right)}
	\text{Tr} \left[ \Gamma 
	\proj_1
	\right]
	\delta \left( \overline{\rvec} - \rvec^\prime\right)
	},
\end{eqnarray}
where $V_N$ is the ring polymer potential, $\Gamma$ is the nuclear-electronic state
interaction term, both previously defined in Eqs.~\eqref{eq:rppot} and \eqref{eq:gamma}, 
and the projection operator
$\proj_1 = \ket{1}\bra{1}$ projects the $N^{th}$ electronic ring polymer bead
onto state 1. 

\begin{figure}[h]
	\centering
 	 \includegraphics[height=3cm]{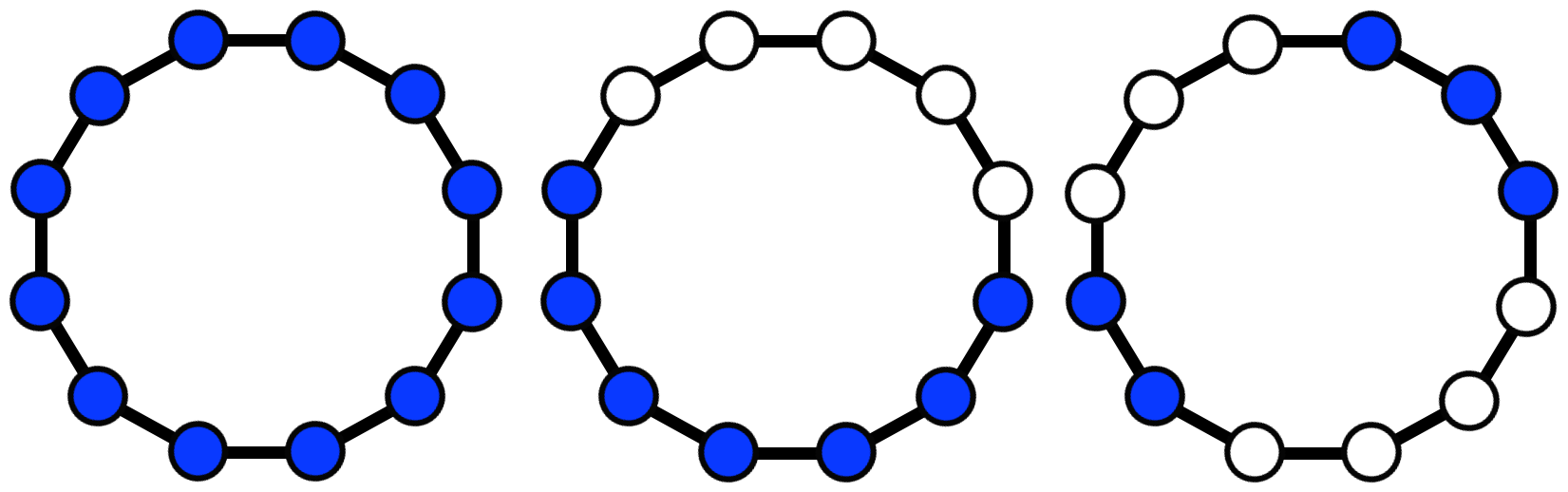}
 	 \caption{An illustration of ring polymer configurations with zero (left),
	 one (center), and two (right) kink-pairs in a two-state system.  The colors
	 blue and white represent the two states of the system.}
	 \label{fgr:kinks}
\end{figure}

The full $\Gamma$ term in Eq.~\eqref{eq:gamma}
accounts for all possible electronic state configurations; 
these include ring polymer configurations for which 
all electronic beads are in the same state as well as ``kinked" configurations where
at least one bead is in a different electronic state than the others.
The term $\Gamma_\text{Kinks}$ in the numerator of Eq.~\eqref{eq:prdagndag}
refers to the subset of $\Gamma$ that
includes only these kinked configurations,
and 
the term $\Gamma \proj_1$ in the denominator
accounts for ring polymer configurations where all beads are in the reactant
electronic state as well as kinked configurations. 
As illustrated in Fig.~\ref{fgr:kinks}, 
the cyclicity of the ring polymer ensures that kinks appear in pairs, 
so the phrase ``kink-pairs" is often used when describing these types 
of configurations.  Physically, kink-pair configurations represent 
tunneling states or \textit{instantons},
and their thermal weight is greatest for nuclear configurations at 
which diabatic potentials are 
degenerate.\cite{Chandler1981,Kuki1987,Marchi1991,Ceperley1995,
Richardson2011}

Restraining individual electronic ring polymer beads to a particular state
space configuration is accomplished in the SS-PI framework by inserting 
appropriate projection matrices $\proj_{n_\alpha}$
between the \textbf{M} matrices in Eq.~\eqref{eq:gamma},
where the subscript $n_\alpha$ in $\proj_{n_\alpha}$ refers to the state onto 
which we project the $\alpha^{th}$ bead:
\begin{equation}
	\label{eq:gammaproj}
	\Gamma_{\left\{ n_\alpha\right\}}= \Gamma_{n_1, n_2,\ldots, n_N} = 
	\prod_{\alpha=1}^{N}\textbf{M} (\rvec_\alpha) \proj_{n_\alpha} .
\end{equation}
We then define $\Gamma_\text{Kinks}$ as the sum over all possible combinations 
of ${\left\{n_\alpha\right\}}$ that correspond to kinked configurations. 
The specifics of generating these 
configurations and calculating the quantities in 
Eq.~\eqref{eq:prdagndag} as well as the dynamic recrossing 
factor are described in the implementation details section. 

\subsection{The Population Coordinate}
Defining the TS for ET in
terms of the solvent position with a weak constraint on 
allowed electronic state configurations is typically 
insufficient to describe ET models with high asymmetry 
(near activationless through inverted regimes of Marcus theory).
To overcome this challenge, we define a TS that enforces equal 
populations in the electronic 
states at solvent configurations where 
the reactant and product electronic state energies are equal.
The corresponding MF-RPMD dynamic recrossing factor in Eq.~\eqref{eq:kappa} 
is then computed for a normalized population-based reaction 
coordinate,
\begin{equation}
	\xi \equiv \Delta \proj= 
	\frac{\text{Tr}[\Gamma \proj_2]-\text{Tr}[\Gamma \proj_1]}
	{\text{Tr}[\Gamma \proj_2]+\text{Tr}[\Gamma \proj_1]},
	\label{eq:deltap}
\end{equation}
that distinguishes between 
reactant, TS, and product configurations in all regimes of ET:
\begin{equation}
	\Delta \proj=\left\{
	\begin{array}{cl}
		-1 & \text{reactant minimum}\\
		0 & \text{transition state}\\
		1 & \text{product minimum}.\\
	\end{array}\right. 
	\label{eq:dproj_vals}
\end{equation}

For the population coordinate, the probability of reaching the TS 
from the reactant state in Eq.\eqref{eq:ktst} is defined as
\begin{eqnarray}
	\label{eq:prdagndagpop2}
	\text{P} \left( \rvec^\ddag, n^\ddag \right) =
	\frac{
	\int \left\{d\rvec_\alpha \right\} 
	e^{-\frac{\beta}{N} 
	V_N
	\left( \{\rvec_\alpha\}\right)}
	\text{Tr} \left[ \Gamma_{\Delta \proj = 0}
	\right]
	\delta \left( \overline{\rvec} - \rvec^\ddag \right)
	}
	{
	\int \left\{d\rvec_\alpha \right\} 
	e^{-\frac{\beta}{N} 
	V_N
	\left( \{\rvec_\alpha\}\right)}
	\text{Tr} \left[ \Gamma_{\Delta \proj = -1}
	\right]
	},
\end{eqnarray}
where  $\Gamma_{\Delta \proj = 0}$ includes only kinked configurations with 
an equal number of ring polymer beads in either state,
$\Gamma_{\Delta \proj = -1}$ includes only configurations where 
all the ring polymer beads are in the reactant state,
and, in the numerator, the nuclear COM is restrained to the 
position at which the two electronic states are degenerate.

\section{Implementation Details}
\subsection{Solvent Reaction Coordinate}
In practice, it is easiest to evaluate the probability
of forming configurations corresponding to the TS 
in Eq.~\eqref{eq:prdagndag} by splitting it into two terms:
\begin{eqnarray}
	\label{eq:psplit}
	\text{P} \left( \rvec^\ddag, n^\ddag \right) = 
	\text{P} \left( \rvec^\ddag \right)
	\times
	\text{P} \left( n^\ddag | \rvec^\ddag \right),
\end{eqnarray}
where
\begin{eqnarray}
	\label{eq:prdag}
	\text{P} \left( \rvec^\ddag \right) = 
	\frac{
	\int \left\{d\rvec_\alpha \right\} 
	e^{-\frac{\beta}{N} 
	V_N
	\left( \{\rvec_\alpha\}\right)}
	\text{Tr} \left[ \Gamma \proj_1
	\right]
	\delta \left( \overline{\rvec} - \rvec^\ddag \right)
	}
	{
	\int_{-\infty}^{\rvec^\ddag} d\rvec^\prime
	\int \left\{d\rvec_\alpha \right\} 
	e^{-\frac{\beta}{N} 
	V_N
	\left( \{\rvec_\alpha\}\right)}
	\text{Tr} \left[ \Gamma 
	\proj_1
	\right]
	\delta \left( \overline{\rvec} - \rvec^\prime\right)
	}
\end{eqnarray}
represents the probability of the system reaching the nuclear TS, 
$\overline{\rvec} = \rvec^\ddag$,
from reactants and
\begin{eqnarray}
	\label{eq:pndag}
	\text{P} \left( n^\ddag | \rvec^\ddag \right) = 
	\frac{
	\int \left\{d\rvec_\alpha \right\} 
	e^{-\frac{\beta}{N} 
	V_N
	\left( \{\rvec_\alpha\}\right)}
	\text{Tr} \left[ \Gamma_\text{Kinks}
	\right]
	\delta \left( \overline{\rvec} - \rvec^\ddag \right)
	}
	{
	\int \left\{d\rvec_\alpha \right\} 
	e^{-\frac{\beta}{N} 
	V_N
	\left( \{\rvec_\alpha\}\right)}
	\text{Tr} \left[ \Gamma 
	\proj_1
	\right]
	\delta \left( \overline{\rvec} - \rvec^\ddag \right)
	}
\end{eqnarray}
represents the conditional probability of the system forming the electronic TS (kink-pair configurations)
given that the solvent COM is at $\rvec^\ddag$.

We evaluate Eq.~\eqref{eq:prdag} by 
generating a free energy profile along
$\overline{\rvec}$
using umbrella sampling\cite{Torrie1977} and the weighted
histogram analysis method (WHAM),\cite{Kumar1992}
where a harmonic restraint on $\overline{\rvec}$ is used to center simulation windows at
different values $\rvec_i$ throughout the reactant and TS regions.
In each window, nuclear configurations are generated
by MC importance sampling using the weighting function 
\begin{equation}
	W_1 = 
	e^{-\frac{\beta}{N} V_N - 0.5 k_c \left( \overline{\rvec} - \rvec_i \right)^2} 
	\text{Tr} \left[ \Gamma \proj_1 \right].
	\label{eq:feweight}
\end{equation}

In a separate simulation, Eq.~\eqref{eq:pndag} is evaluated 
by importance sampling using the weighting function
\begin{equation}
	W_2 = 
	e^{-\frac{\beta}{N} V_N},
	\label{eq:kinkweight}
\end{equation}
and the delta function $\delta \left( \overline{\rvec} - \rvec^\ddag \right)$ 
is enforced by shifting the nuclear ring polymer COM to $\rvec^\ddag$ 
for each MC step.
The terms $\text{Tr} \left[ \Gamma_\text{Kinks} \right]$ and 
$\text{Tr} \left[ \Gamma \proj_1 \right]$ are evaluated at 
each step, and the ratio of their final
averages yields $\text{P} \left( n^\ddag | \rvec^\ddag \right)$.

In order to calculate $\text{Tr} \left[ \Gamma_\text{Kinks} \right]$,
we must account for all combinations of the set
$\left\{ n_\alpha \right\}$ in Eq.~\eqref{eq:gammaproj}
that correspond to kinked configurations.
Consider a two-state system ($K = 2$) for simplicity.
A particular electronic configuration 
$\left\{ n_\alpha \right\}~\equiv~\left\{j,w,m\right\}$ is characterized 
by the number of beads in state 1 which we denote as $j$,
the number of kink-pairs present which we denote as $w$,
and $m$ which represents the particular electronic configuration 
in the subset of configurations that have the same values of $j$ and $w$.
Combinations for which there exist at least one kink-pair
correspond to values of $j$ equal to $1$ through $N - 1$, 
and the number of possible kink-pairs
ranges from 1 to $w_\text{tot}$, where $w_\text{tot} = j$ for $j \le N/2$ and 
$w_\text{tot} = N - j$ for $j > N/2$;
values for $m$ range from 1 to $m_\text{tot}$, where $m_\text{tot}$ 
depends on the particular values of $j$ and $w$.
For a given nuclear configuration the exact thermal weight of kink-pair 
configurations is
\begin{equation}
	\label{eq:gammakinks}
	\text{Tr} \left[ \Gamma_{\text{Kinks}} \right] = 
	\text{Tr} \left[
	\sum_{j =1}^{N-1}
	\sum_{w =1}^{w_\text{tot}\left(j\right)}
	\sum_{m=1}^{m_\text{tot}\left(j,w\right)}
	\Gamma_{\left\{j,w,m\right\}} \right].
\end{equation}

For a large number of ring polymer beads, 
we acheive an efficient implementation by evaluating 
Eq.~\eqref{eq:gammakinks} once at the beginning of the 
simulation to determine $m_\text{tot}\left(j,w\right)$.
We then choose a ``representative configuration" 
$\left\{ n_\alpha \right\}~\equiv~\left\{j,w\right\}^\prime$ for
every combination of $j$ and $w$.
This allows us to evaluate
$\text{Tr} \left[ \Gamma_\text{Kinks} \right]$
at each MC step as a sum over representative combinations
weighted by $m_\text{tot}\left(j,w\right)$,
\begin{equation}
	\label{eq:gammakinks2}
	\text{Tr} \left[ \Gamma_{\text{Kinks}} \right] = 
	\text{Tr} \left[
	\sum_{j =1}^{N-1}
	\sum_{w =1}^{w_\text{tot}\left(j\right)}
	m_\text{tot}\left(j, w\right)
	\Gamma_{\left\{j, w\right\}^\prime} \right],
\end{equation}
which, on average, yields the same result as Eq.~\eqref{eq:gammakinks}.
In the weak coupling regime, sampling can be limited 
to configurations with $w=1$ that dominate the sum;
however, in the present work we do not find it necessary to
impose this limit on the number of kink-pairs.
Finally, in order to evaluate $\text{Tr}\left[ \Gamma\proj_1 \right]$ in 
Eq.~\eqref{eq:pndag} for a given nuclear configuration, 
we simply add to $\text{Tr}\left[ \Gamma_\text{Kinks} \right]$ 
a term that corresponds to all the RP beads being in electronic state 1 
($j=N$ and $w=0$).

The average forward velocity term
that appears in the numerator of the TST estimate
and the denominator of the dynamic recrossing factor 
can be analytically obtained by evaluating a Gaussian integral 
in the solvent momentum:
\begin{equation}
	\left\langle 
	\dot{\overline{\rvec}}_0 \; h\left(\dot{\overline{\rvec}}_0 \right) \right\rangle_c = 
	\left( \frac{1}{2 \pi \beta M} \right)^{d/2}.
	\label{eq:forwardvel}
\end{equation}

Initial configurations for MF-RPMD trajectories 
are generated by importance sampling using the weighting
function
\begin{equation}
	W_3 = 
	e^{-\frac{\beta}{N} V_N}
	\text{Tr} \left[ \Gamma_\text{Kinks} \right],
	\label{eq:mfrpmdweight}
\end{equation}
and the delta function $\delta\left( \overline{\rvec}-\rvec^\ddag \right)$
is enforced by shifting the nuclear COM to $\rvec^\ddag$ after each MC step.
Here, the term $\text{Tr} \left[ \Gamma_\text{Kinks} \right]$
is evaluated using Eq.~\eqref{eq:gammakinks2}.
MF-RPMD trajectories are evolved in time using the 
classical ring polymer Hamiltonian in Eq.~\eqref{eq:mfham};
averaging the expression
$\left( \dot{\overline{\rvec}}_0 \; h \left(\overline{\rvec}_t - \rvec^\ddag \right) \right)$ over all trajectories and
dividing by Eq.~\eqref{eq:forwardvel} yields a value 
for $\kappa_\text{MF-RPMD}$.

\subsection{Population Reaction Coordinate}
As with the solvent reaction coordinate, 
we evaluate the probability of forming configurations 
corresponding to the population coordinate TS in 
Eq.~\eqref{eq:prdagndagpop2} by splitting it into two terms,
where
\begin{eqnarray}
	\label{eq:prdagpop}
	\text{P} \left( \rvec^\ddag \right) = 
	\frac{
	\int \left\{d\rvec_\alpha \right\} 
	e^{-\frac{\beta}{N} 
	V_N
	\left( \{\rvec_\alpha\}\right)}
	\text{Tr} \left[ \Gamma_{\Delta \proj = -1}
	\right]
	\delta \left( \overline{\rvec} - \rvec^\ddag \right)
	}
	{
	\int \left\{d\rvec_\alpha \right\} 
	e^{-\frac{\beta}{N} 
	V_N
	\left( \{\rvec_\alpha\}\right)}
	\text{Tr} \left[ \Gamma_{\Delta \proj = -1}	\right],
	}
\end{eqnarray}
and
\begin{eqnarray}
	\label{eq:pndagpop}
	\text{P} \left( n^\ddag | \rvec^\ddag \right) = 
	\frac{
	\int \left\{d\rvec_\alpha \right\} 
	e^{-\frac{\beta}{N} 
	V_N
	\left( \{\rvec_\alpha\}\right)}
	\text{Tr} \left[ \Gamma_{\Delta \proj = 0}
	\right]
	\delta \left( \overline{\rvec} - \rvec^\ddag \right)
	}
	{
	\int \left\{d\rvec_\alpha \right\} 
	e^{-\frac{\beta}{N} 
	V_N
	\left( \{\rvec_\alpha\}\right)}
	\text{Tr} \left[ \Gamma_{\Delta \proj = -1}
	\right]
	\delta \left( \overline{\rvec} - \rvec^\ddag \right)
	}.
\end{eqnarray}
Eq.~\eqref{eq:prdagpop} is evaluated with the same techniques 
used for Eq.~\eqref{eq:prdag},
but here we employ the weighting function
\begin{equation}
	W_4 = 
	e^{-\frac{\beta}{N} V_N - 0.5 k_c \left( \overline{\rvec} - \rvec_i \right)^2} 
	\text{Tr} \left[ \Gamma_{\Delta \proj = -1} \right],
	\label{eq:feweight}
\end{equation}
where
\begin{equation}
	\label{eq:gammadpreact}
	\text{Tr} \left[ \Gamma_{\Delta \proj = -1} \right] = 
	\text{Tr} \left[ \Gamma_{\left\{N,0\right\}^\prime} \right].
\end{equation}
We evaluate Eq.~\eqref{eq:pndagpop} using an approach similar to Eq.~\eqref{eq:pndag},
but in this case 
we only include kinked configurations with equal numbers
of beads in each state:
\begin{equation}
	\label{eq:gammadp0}
	\text{Tr} \left[ \Gamma_{\Delta \proj = 0} \right] = 
	\text{Tr} \left[
	\sum_{w =1}^{N/2}
	m_\text{tot}\left(N/2, w\right)
	\Gamma_{\left\{N/2, w\right\}^\prime} \right].
\end{equation}

Initial configurations for the MF-RPMD simulation are 
generated by importance sampling using the weighting function
\begin{equation}
	W_5 = 
	e^{-\frac{\beta}{N} V_N}
	\text{Tr} \left[ \Gamma_{\Delta \proj = 0} \right],
	\label{eq:mfrpmdweight}
\end{equation}
and again we implement 
$\delta\left( \overline{\rvec}-\rvec^\ddag \right)$ by shifting the 
nuclear COM to $\rvec^\ddag$.
Trajectories initially constrained to this TS distribution are evolved 
using the Hamiltonian in Eq.~\eqref{eq:mfham}, and the average initial 
velocity of the population coordinate is determined by computing the 
rate of change of $\Delta\proj$ for each trajectory using a finite 
difference approximation at very short times and averaging over the ensemble. 
The average forward velocity computed using this technique 
is then multiplied by $\text{P} \left(\rvec^\ddag, n^\ddag \right)$ to obtain
the TST rate estimate. 

\subsection{Rate Theories for Adiabatic and Nonadiabatic Electron Transfer}
The Marcus theory (MT) rate for a nonadiabatic ET reaction 
with a classical solvent is
\cite{Marcus1985}
\begin{equation}
k_{\text{MT}} = 
\frac{2 \pi}{\hbar} {| \Delta|}^2 
\sqrt{\frac{\beta}{4 \pi \lambda}} 
e^{-\beta \left( \lambda - \varepsilon \right)^2 / 4 \lambda},
\label{eq:kmt}
\end{equation}
where $\lambda$ is the solvent reorganization energy, 
$\varepsilon$ is the
asymmetry between the reactant and product state energies at their respective minima,
and $\Delta$ is the coupling between the reactant and product diabatic 
electronic states.

The nonadiabatic ET rate with a quantized solvent 
can be calculated using Fermi's golden rule (FGR). 
For systems in which the reactant and product
diabatic potential energy surfaces are displaced harmonic 
oscillators with frequency $\omega_s$, FGR rates 
take the simple analytical form\cite{Ulstrup1975,Ulstrup1979}
\begin{equation}
k_{\text{FGR}} = 
\frac{2 \pi}{\hbar \omega_s} {| \Delta|}^2 
e^{v z - S \; \text{coth}\left( z \right)} I_v \left(S \; \text{csch}\left(z\right)\right),
\label{eq:kfgr}
\end{equation}
where $z = \beta \omega_s / 2$, $v = \varepsilon/\omega_s$, 
$S = M_s \omega_s V_d^2 / 2 \hbar$,
$M_S$ is the solvent mass,
$I_v$ is a modified Bessel function of the first kind, and
$V_d$ is the horizontal displacement of the diabatic potential energy functions.

Reaction rates for ET in the adiabatic limit with a quantum solvent 
can be estimated using Kramers theory (KT),
\cite{Henriksen2008}
\begin{equation}
k_{\text{KT}} = 
\left(
\sqrt{ 1 + \left( \frac{\gamma}{2 \omega_b} \right)^2 }
- \frac{\gamma}{2 \omega_b} \right)
\frac{\omega_s}{2 \pi} 
e^{-\beta G^\ddag_\text{cl}},
\label{eq:kkt}
\end{equation}
where $\omega_b$ is the frequency that confines the
barrier top, $G^\ddag_\text{cl}$ is the solvent FE barrier when the solvent
is treated classically, $\gamma~=~\eta / M_S,$\cite{Makri1996}
and $\eta$ is the strength of coupling to a dissipative bath.

\section{Model Systems}
\begin{table}[h]
\small
\centering
  \caption{~Parameters for ET models.  Unless otherwise specified, values are reported
  in atomic units.}
  \label{tbl:modelParams}
  \begin{tabular}{cc} 
    \hline
    Parameter & Value Range\\
    \hline
    $A$ & $4.772 \times 10^{-3}$ \\
    $B$ & $2.288 \times 10^{-2}$ \\
    $\varepsilon$ & 0 - 0.2366 \\
     $\Delta$ & $6.69\times 10^{-7}$ - $1.20\times 10^{-2}$ \\
    $M_S$ & 1836.0 \\
    $M_B$ & 1836.0 \\
    $f$ & 12 \\
    $\omega_c$ & $2.28 \times 10^{-3}$ \\
    $\eta / M \omega_c$ & 1.0 \\
    $T$ & 300 K \\
    \hline
  \end{tabular}
\end{table}
Numerical results are presented for condensed-phase ET systems 
with potential energy functions of the form\cite{Topaler1994,Menzeleev2014}
\begin{equation}
	V \left( \hat{\rvec} \right) = V_S \left( \hat{\rvec} \right) + 
	\mathbb{1} V_B \left( \hat{\rvec} \right),
	\label{eq:potential}
\end{equation}
where $\mathbb{1}$ is the identity matrix and $\rvec = \left\{s,\Qvec\right\}$ represents the full set of 
nuclear coordinates, including both a solvent polarization coordinate, $s$, and bath coordinates, $\Qvec$.  
The diabatic potential energy matrix for each system, constructed along the solvent
coordinate, has the form
\begin{equation}
	V_S \left( \hat{\rvec} \right) = \begin{bmatrix}
	A \hat{s}^2 + B \hat{s}  + \varepsilon & \Delta \\ 
	\Delta & A \hat{s}^2 - B \hat{s}
	\end{bmatrix}
	\label{eq:diabatmatrix},
\end{equation}
and the solvent coordinate, with associated mass $M_S$,
is linearly coupled to a set of $f$ harmonic oscillators, with mass $M_B$,
through the potential
\begin{equation}
	V_B \left( \hat{\rvec} \right) = \sum_{j =1}^f \left[ \frac{1}{2} M_B \omega^2_j 
	\left(\hat{Q}_j - \frac{c_j \hat{s}}{M_B \omega^2_j} \right)^2 \right].
	\label{eq:bathpotential}
\end{equation}
The spectral density of the bath is Ohmic, 
\begin{equation}
	J \left( \omega \right) = 
	\eta \omega e^{-\omega / \omega_c}
	\label{eq:ohmic},
\end{equation}
with cutoff frequency $\omega_c$ and dimensionless parameter $\eta$ that determines the
friction strength of the bath.
Following the scheme developed in Ref.~\citenum{Craig20052}, we discretize the spectral density 
into $f$ oscillators with frequencies
\begin{equation}
	\omega_j = - \omega_c \; \text{ln} \left( \frac{j - 1/2}{f} \right)
	\label{eq:freqs}
\end{equation}
and coupling strengths 
\begin{equation}
	c_j = \omega_j \left(\frac{2 \eta M_B \omega_c}{f \pi} \right)^{1/2}
	\label{eq:coups},
\end{equation}
where $j = 1...f$.
We test a range of driving force values, $\varepsilon$,
as well as a range of electronic coupling strengths, $\Delta$,
from the nonadiabatic to adiabatic limit.
In all cases considered, we quantize all degrees of freedom with $N = 32$ ring polymer beads.  
All other parameters are reported in Table~\ref{tbl:modelParams}.

\section{Results and Discussion}
We calculate nonadiabatic reaction rates 
for the model ET systems described in Section 5
over a wide range of driving forces, electronic
coupling strengths, and with different reaction coordinates. 

\begin{figure}[!h]
	\centering
 	 \includegraphics[scale=0.55]{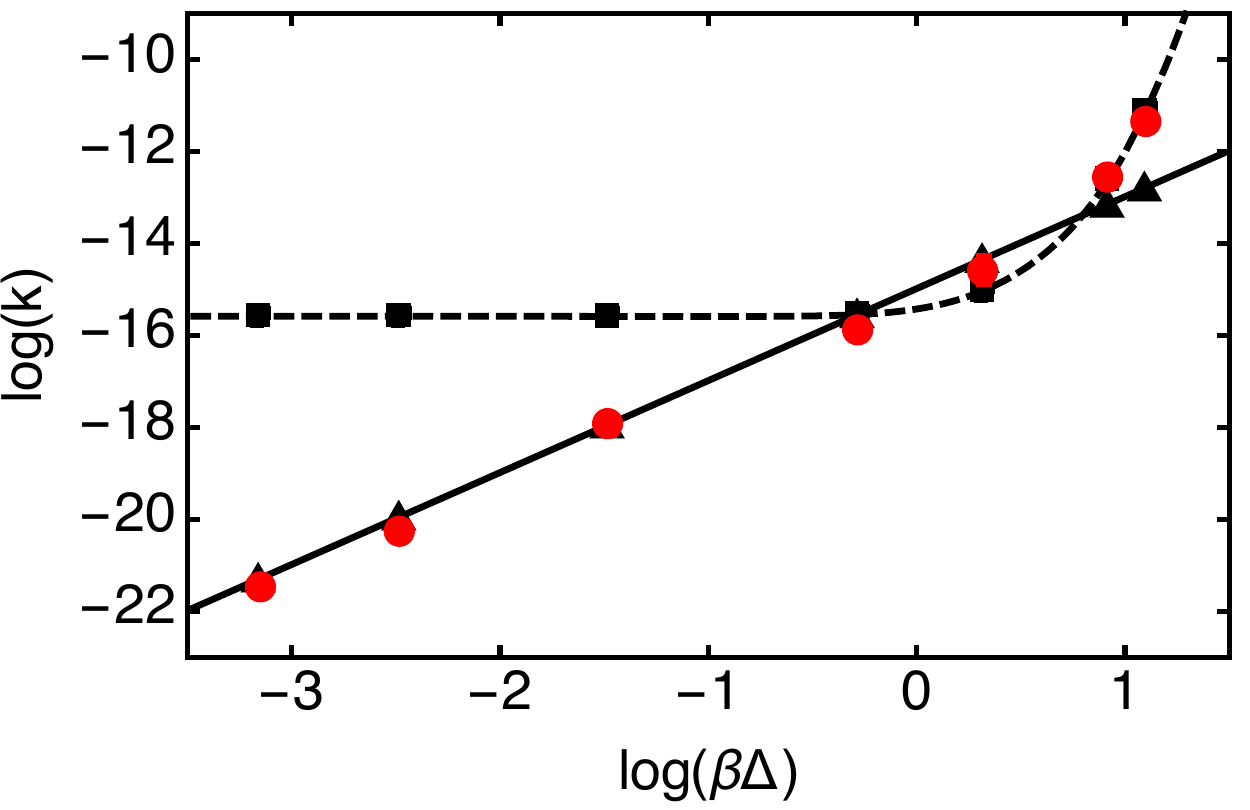}
 	 \caption{ET rate constants computed using the solvent
	 reaction coordinate for a range of electronic coupling constants, $\Delta$,
	 for the symmetric model, $\varepsilon = 0$.  MF-RPMD
	 values are shown in red dots, FGR rate constants are shown in
	 black triangles and a solid black line, and Kramers theory
	 rate constants are shown in black squares and a black dashed line.
	 Both axes are in atomic units.}
	 \label{fgr:coupling}
\end{figure}
First, we present our results for ET reaction rates using the solvent reaction coordinate
for the symmetric case, Model I ($\varepsilon = 0$), with different 
electronic coupling values. 
For all calculations that employ the solvent reaction coordinate, 
TST results are obtained using a force constant 
$k_c = 200$ a.u.~in umbrella sampling,
and MF-RPMD results are obtained by averaging over
$24,000$ trajectories evolved using a time step $dt = 0.1$ a.u.
In Fig.~\ref{fgr:coupling}, we compare our results against 
the Kramers theory rates for adiabatic ET and FGR for nonadiabatic ET, and
we show that our calculated rates
exhibit quantitative agreement with the 
applicable theory across six orders of magnitude 
in the electronic coupling.
Numerical values for the rate constants are reported in 
Table~\ref{tbl:delta_range}, along 
with the TST rate. We see that, despite the limitations of 
MF-RPMD, the accuracy of the TST rate in this regime is sufficient
for good numerical agreement.
\begin{table}[!htb]
\small
\centering
  \caption{~ET rates for a range of electronic coupling strengths, $\Delta$, for the symmetric
  model, $\varepsilon = 0$, computed using the solvent reaction coordinate.  
  From left to right, the four rightmost columns
  report the TST estimate to the rate constant, the full MF-RPMD rate constant,
  Fermi's golden rule values, and Kramers theory rate constants, respectively.
  The numbers in parentheses represent the statistical uncertainty in the last
  digit reported, and
  all values are reported in atomic units.}
  \label{tbl:delta_range}
  \begin{tabular}{ccccc} 
    \hline
    $\Delta$ & 
    $\text{log} \left(k_\text{TST} \right)$ & 
    $\text{log} \left(k_\text{MF-RPMD} \right)$ &
    $\text{log} \left(k_\text{FGR} \right)$ &
    $\text{log} \left(k_\text{KT} \right)$  \\
    \hline
     $6.69\times 10^{-7}$ & -21.47 & -21.47(8) & -21.28 & -15.57\\
     $3.16\times 10^{-6}$ & -20.22 & -20.2(2) & -19.93 & -15.58\\
     $3.16\times 10^{-5}$ & -17.95 & -17.9(2) & -17.93 & -15.58\\
     $5.01\times 10^{-4}$ & -15.84 & -15.8(1) & -15.53 & -15.54\\
     $2.00\times 10^{-3}$ & -14.55 & -14.6(3) & -14.33 & -15.02\\
     $7.94\times 10^{-3}$ & -12.51 & -12.55(4) & -13.13 & -12.60\\
     $1.20\times 10^{-2}$ & -11.30 & -11.3(2) & -12.77 & -11.11\\
    \hline
  \end{tabular}
\end{table}

\begin{figure}[!htb]
	\centering
 	 \includegraphics[scale=0.55]{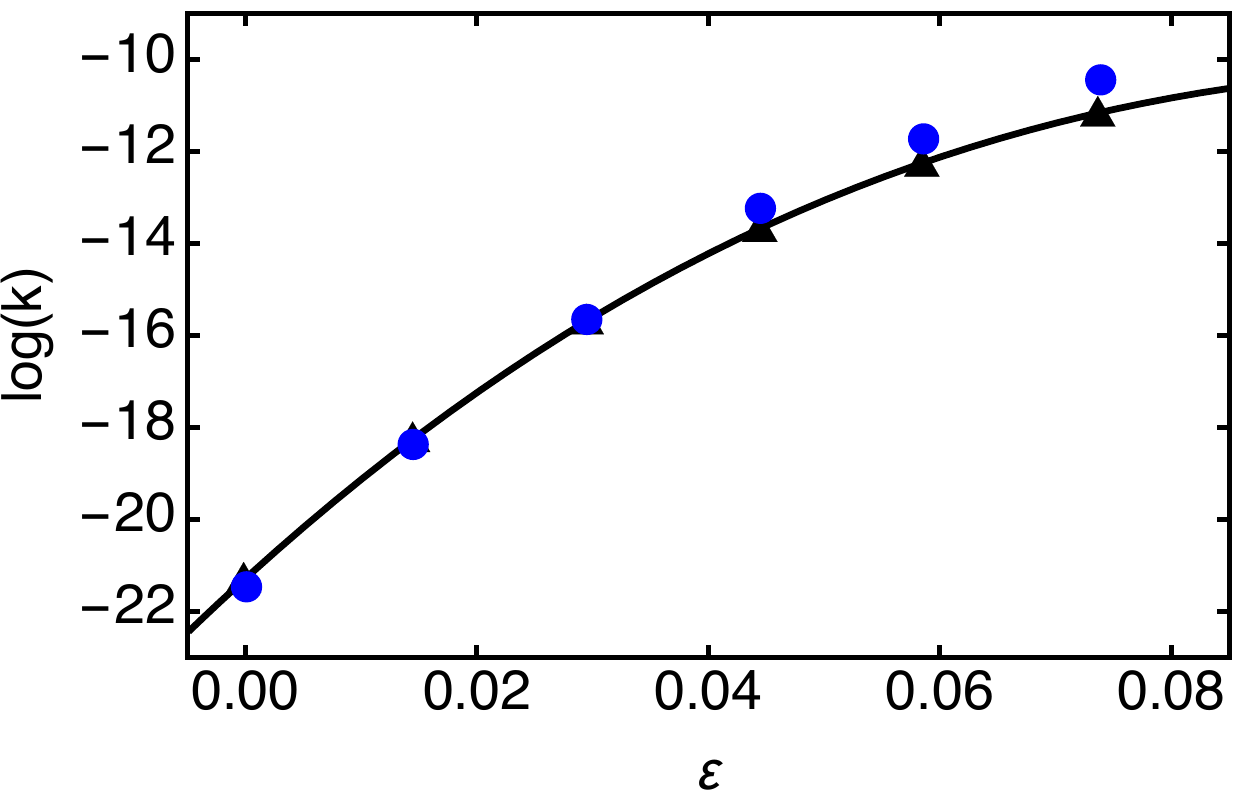}
 	 \caption{ET rate constants computed using the
	 solvent reaction coordinate for a range of driving force values, $\varepsilon$,
	 with constant coupling $\Delta = 6.69 \times 10^{-7}$.  MF-RPMD results
	 are shown in blue dots, and FGR rate constants are shown in
	 black triangles and a solid black line.
	 Both axes are in atomic units.}
	 \label{fgr:drivingforcesol}
\end{figure}

\begin{figure}[!htb]
	\centering
 	 \includegraphics[scale=0.55]{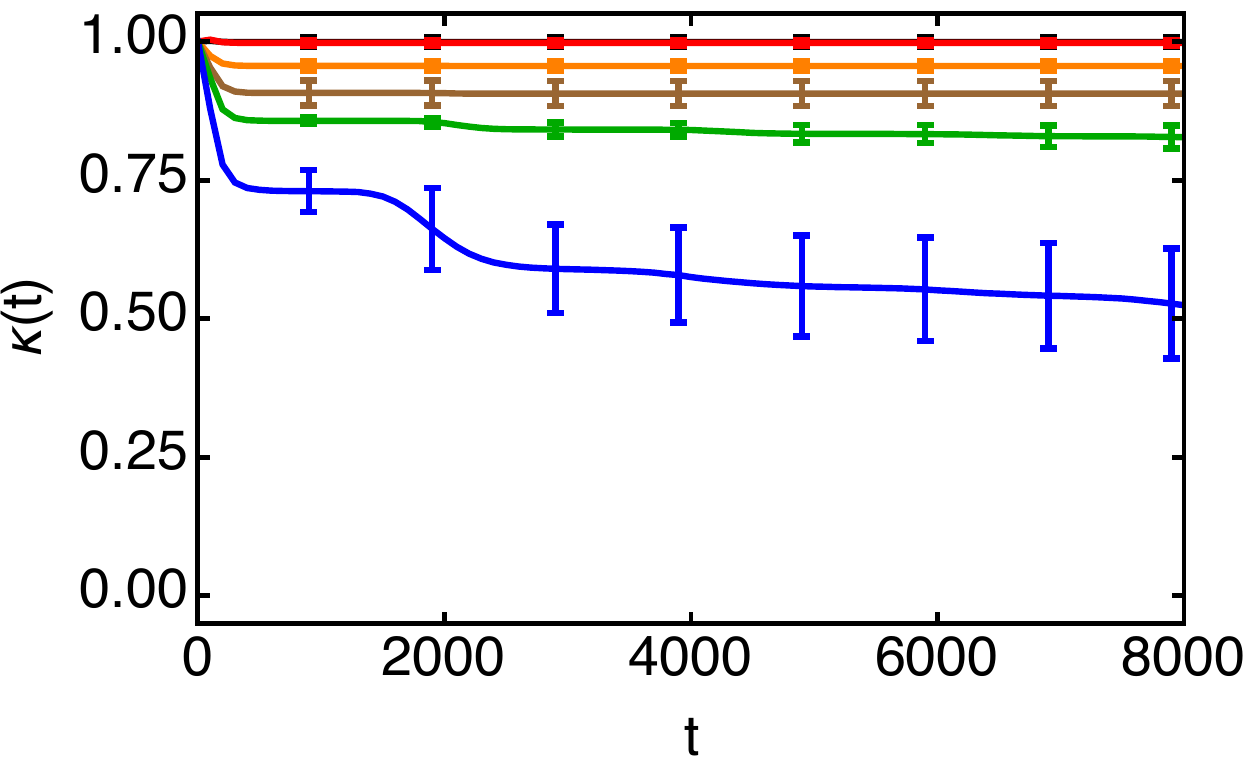}
 	 \caption{Plots of the dynamical recrossing term, $\kappa_\text{MF-RPMD}(t)$, 
	 computed using the solvent reaction coordinate as a function of time
	 for Models I-VI (black, red, orange, brown, green, and blue, respectively) from top to bottom.
	 Both axes are in atomic units.}
	 \label{fgr:kappasol}
\end{figure}

Next, we present the rate of ET calculated 
using the solvent coordinate for weak-coupling Models I-VI
that explore a range of driving forces in the normal regime.
Fig.~\ref{fgr:drivingforcesol} compares our MF-RPMD rates 
to FGR rates.
The exact values of these rate constants
are also reported in Table~\ref{tbl:epsilonlistsol}, along with 
the state space TST estimates.  
We also show the dynamic recrossing factor as a function of 
time in Fig.~\ref{fgr:kappasol}.

We find that our MF-RPMD implementation proves quantitatively accurate
for ET in the normal regime. The high values of $\kappa$, particularly
for the symmetric and near-symmetric models of ET, 
demonstrate the accuracy of our TST rate for these models.
As the models become more asymmetric, $\kappa_\text{MF-RPMD}$ decreases,
and eventually, as seen in Model VI (blue curve), at longer times we no 
longer observe plateau behavior (we use the value of $\kappa_\text{MF-RPMD}$
at $t = 8000$ a.u.~to obtain the reported rate constant for this model).
MF-RPMD with the solvent reaction coordinate
becomes inapplicable beyond Model VI--this is expected since 
the solvent coordinate is no longer able to distinguish between 
reactant and product states.

\begin{figure}[!h]
	\centering
 	 \includegraphics[scale=0.55]{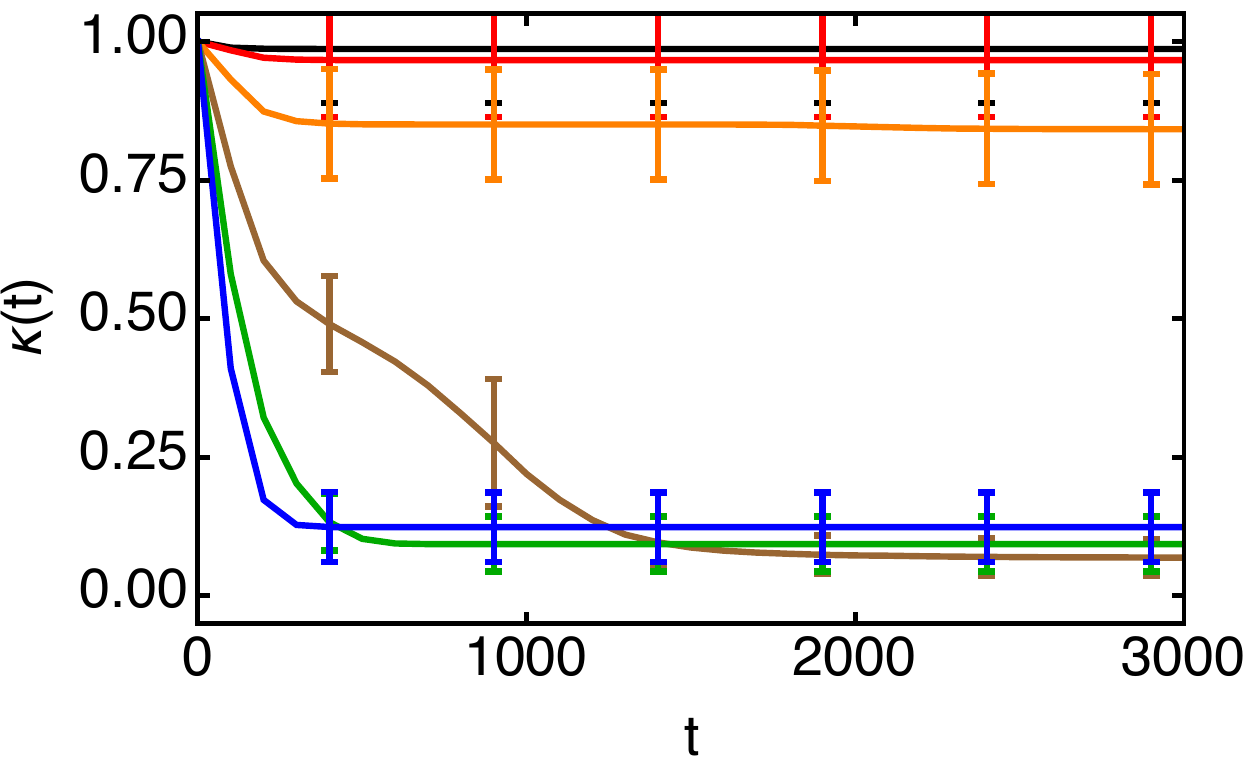}
 	 \caption{Plots of the dynamical recrossing term, $\kappa_\text{MF-RPMD}(t)$, 
	 computed using the population reaction coordinate as a function of time
	 for Models I, III, V, and VII-IX 
	 (black, red, orange, brown, green, and blue, respectively).
	 Both axes are in atomic units.}
	 \label{fgr:kappapop}
\end{figure}

\begin{table}[h]
\small
\centering
  \caption{~ET rates computed using the solvent reaction coordinate 
  for a range of driving forces, $\varepsilon$, with
  constant coupling $\Delta = 6.69 \times 10^{-7}$.  
  From left to right, the three rightmost columns
  report the TST estimate to the rate constant, 
  the full MF-RPMD rate constant, and
  the Fermi's golden rule values, respectively.
  The numbers in parentheses represent the statistical 
  uncertainty in the last digit reported, and
  all values are reported in atomic units.
  }
  \label{tbl:epsilonlistsol}
  \begin{tabular}{ccccc} 
    \hline
    Model & $\varepsilon$ &
    $\text{log} \left(k_\text{TST} \right)$ & 
    $\text{log} \left(k_\text{MF-RPMD} \right)$ &
    $\text{log} \left(k_\text{FGR} \right)$  \\
    \hline
    I & 0.0000 & -21.47 & -21.47(8) & -21.28 \\
    II & 0.0146 & -18.35 & -18.349(6) & -18.23 \\
    III & 0.0296 & -15.65 & -15.670(5) & -15.66 \\
    IV & 0.0446 & -13.18 & -13.22(1) & -13.65 \\
    V & 0.0586 & -11.60 & -11.69(1) & -12.23 \\
    VI & 0.0738 & -10.18 & -10.47(8) & -11.15 \\
    \hline
  \end{tabular}
\end{table}
\begin{figure}[!htb]
	\centering
 	 \includegraphics[scale=0.55]{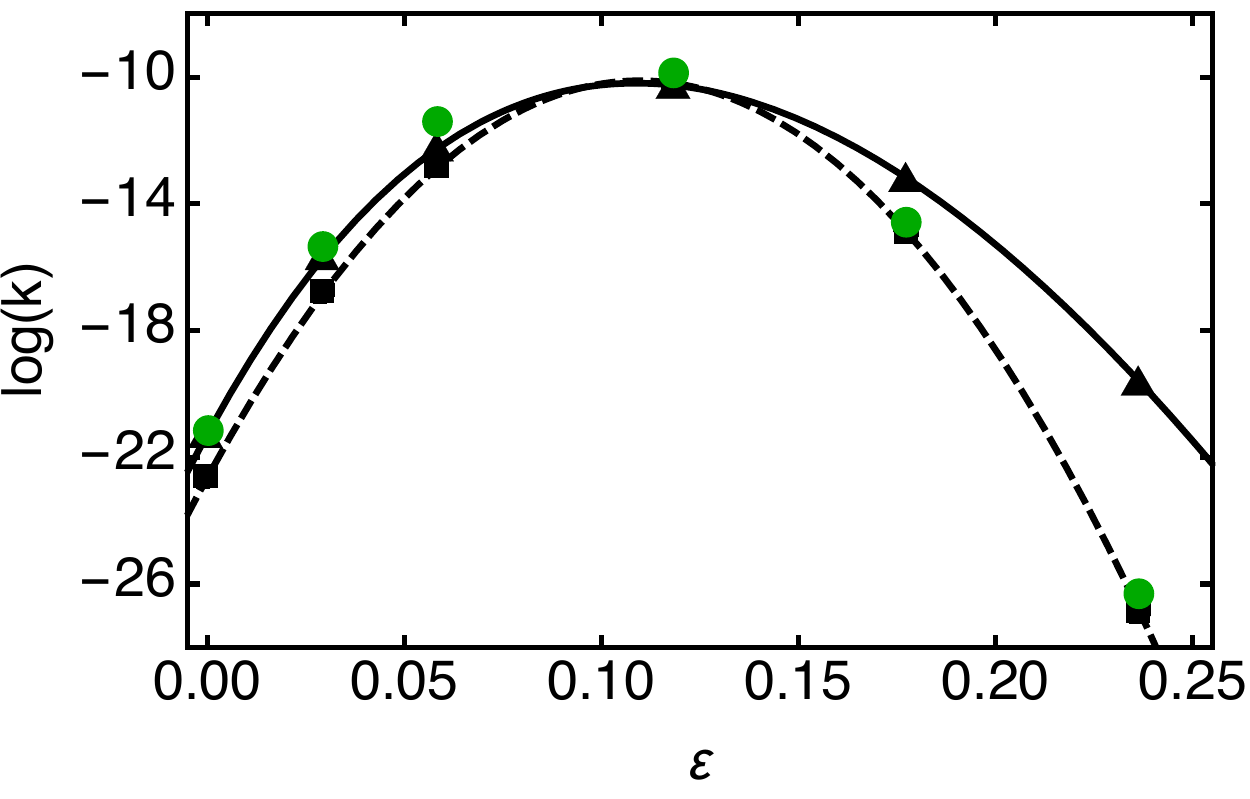}
 	 \caption{ET rate constants computed using the
	 population reaction coordinate for a range of driving force values, $\varepsilon$,
	 with constant coupling $\Delta = 6.69 \times 10^{-7}$.  MF-RPMD results
	 are shown in green dots, FGR rate constants are shown in
	 black triangles and a solid black line, and MT results
	 are shown in black squares and a black dashed line.
	 Both axes are in atomic units.}
	 \label{fgr:drivingforcepop}
\end{figure}
Finally, we present our results for ET rates
calculated using the population coordinate in Models 
I, III, and V (normal regime) and in Models VII-IX (activationless
and inverted regimes).
For these simulations, TST results are obtained using a 
force constant $k_c = 200$ a.u.~in umbrella sampling.
MF-RPMD results are obtained by averaging over
$30,000$ trajectories evolved using a time step $dt = 0.1$ a.u.,~and
numerical derivatives used to compute the 
initial $\Delta \proj$ velocities
are calculated by averaging 
$\left( \Delta \proj \left(n \times dt \right) - 
\Delta \proj \left(0\right) \right)/ 
\left( n \times dt \right)$ for $n = 20, 30,$ and $40$.
Fig.~\ref{fgr:drivingforcepop} shows that rates obtained using 
the population coordinate, like the solvent coordinate, are quantitatively
accurate, agreeing with FGR rates in the normal regime.
Additionally, we are able to move past Model VI to the activationless
and inverted regimes (Models VII-IX), 
where the population coordinate remains a good reaction coordinate.
The numerical values of our calculated rates, along with the TST rates, are 
reported in Table~\ref{tbl:epsilonlistpop}.
Further, Fig.~\ref{fgr:kappapop} shows $\kappa(t)$ for the different models;
as in the previous case, $\kappa$ is approximately 1 for the symmetric
model and decreases as the driving force increases. \\
\indent 
We note that in the inverted regime our results are qualitatively reasonable
and capture the predicted Marcus turnover in rates.  However, we do not find 
quantitative agreement with FGR; instead, our results agree more closely
with Marcus theory rates. We attribute this to the fact that
our definition of $k_\text{TST}$ does not allow 
kinked configurations of the ring polymer to form 
except at solvent configurations corresponding to the point of degeneracy 
between the two diabats.~\cite{Menzeleev2014} We expect that 
either using a better formulation for the TS that can explicitly
account for solvent tunneling or employing dynamics beyond mean field will 
improve our numerical results in the inverted regime.

\begin{table}[!h]
\small
\centering
  \caption{~
  ET rates computed using the population coordinate
  for a range of driving forces, $\varepsilon$, with constant coupling,
  $\Delta = 6.69 \times 10^{-7}$.  From left to right, the four rightmost columns
  report the TST estimate to the rate constant, the full MF-RPMD rate constant,
  Fermi's golden rule values, and Marcus theory rate constants, respectively.
  The numbers in parentheses represent the statistical uncertainty in the last
  digit reported, and
  all values are reported in atomic units.}
  \label{tbl:epsilonlistpop}
  \begin{tabular}{cccccc} 
    \hline
    Model & $\varepsilon$ & 
    $\text{log} \left(k_\text{TST} \right)$ & 
    $\text{log} \left(k_\text{MF-RPMD} \right)$ &
    $\text{log} \left(k_\text{FGR} \right)$ &
    $\text{log} \left(k_\text{MT} \right)$  \\
    \hline
    I & 0.0000 & -21.18(8) & -21.19(9) & -21.28 & -22.65\\
    III & 0.0296 & -15.34(4) & -15.36(6) & -15.66 & -16.79\\
    V & 0.0586 & -11.37(5) & -11.45(7) & -12.23 & -12.83\\
    VII & 0.1186 & -8.72(5) & -9.9(2) & -10.26 & -10.19\\
    VIII & 0.1776 & -13.50(5) & -14.5(2) & -13.20 & -14.91\\
    IX & 0.2366 & -25.44(7) & -26.3(2) & -19.63 & -26.89\\
    \hline
  \end{tabular}
\end{table}

\section{Concluding Remarks}
We show that combining TST rates computed using a state
space path integral formulation with dynamic correction factors computed 
using MF-RPMD yield quantitatively accurate rates for ET over a wide range 
of electronic coupling strengths and driving forces.
This implementation is general for multi-electron, multi-state systems,
and we expect the simple protocol described here to work well for large scale 
atomistic simulations. 
Moving forward, we anticipate that the state space TST implementation 
described here, in combination with nonadiabatic
RPMD methods such as mapping variable (MV)-RPMD,\cite{Ananth2013} 
will prove extremely useful in the study of photo-induced 
charge transfer reactions.


\bibliography{rsc} 
\bibliographystyle{rsc}

\end{document}